\documentclass{PoS}

\usepackage[utf8x]{inputenc}
\usepackage{amsfonts}
\usepackage{amssymb}
\usepackage{amsmath}
\usepackage{amssymb}
\usepackage{graphicx}
\usepackage[]{units}

\let\l\left
\let\r\right
\let\de\partial
\let\mrm\mathrm
\let\mcal\mathcal
\let\mbb\mathbb
\let\dag\dagger

\newcommand{\beq}{\begin{equation}}
\newcommand{\eeq}{\end{equation}}
\newcommand{\beqa}{\begin{eqnarray}}
\newcommand{\eeqa}{\end{eqnarray}}
\newcommand{\ba}{\begin{array}}
\newcommand{\ea}{\end{array}}
\newcommand{\bmat}{\begin{pmatrix}}
\newcommand{\emat}{\end{pmatrix}}
\newcommand{\bcas}{\begin{cases}}
\newcommand{\ecas}{\end{cases}}

\title{Thimble regularization at work besides toy models: from Random Matrix Theory to Gauge Theories.}

\ShortTitle{Thimble regularization from Random Matrix Theory to Gauge Theories.}

\author{\speaker{Giovanni Eruzzi} and Francesco Di Renzo\\
  University of Parma and INFN\\
  E-mail: \email{giovanni.eruzzi@pr.infn.it}, \email{francesco.direnzo@pr.infn.it}
}

\abstract{Thimble regularization as a solution to the sign problem has been successfully put at work for a few toy models. Given the non trivial nature of the method (also from the algorithmic point of view) it is compelling to provide evidence that it works for realistic models. A Chiral Random Matrix theory has been studied in detail. The known analytical solution shows that the model is non-trivial as for the sign problem (in particular, phase quenched results can be very far away from the exact solution). This study gave us the chance to address a couple of key issues: how many thimbles contribute to the solution of a realistic problem? Can one devise algorithms which are robust as for staying on the correct manifold? The obvious step forward consists of applications to gauge theories.}

\FullConference{The 33rd International Symposium on Lattice Field Theory\\
                 14 -18 July 2015\\
                 Kobe International Conference Center, Kobe, Japan}

\begin{document}

\section{Motivation}

When a field theory features complex terms in its action, we say that it is plagued by the so-called sign problem (\emph{e.g.} it usually occurs at finite chemical potential); complex terms prevent numerical simulation of such theories by means of Monte Carlo methods. Several wayouts have been proposed so far, but a universal and rigorous prescription to circumvent the sign problem is still missing. The power of the thimble approach lies in the fact that integrating along the thimble automatically keeps $S_I=\Im\l(S\r)$ constant, thus avoiding the problem of sampling complex actions. A few multi-dimensional models have been successfully investigated with this approach ({\cite{taming,kiku}). Quite interestingly, the simple ones (\emph{e.g.} the 0-dimensional $\phi^4$ toy-model \cite{mytalk14}) showed a non-trivial thimble structure. It is therefore interesting to apply the thimble approach to a simple, yet multi-dimensional model (the CRM theory) to check if more than one thimble is needed to recover the expected results. At the same time, a new method to sample on the thimble will be presented. The details of the following presentation and its application to the CRM theory can be found in \cite{ournew}.

\section{Essential results from Morse theory}

In this section we collect main results from Picard-Lefschetz (\emph{i.e.} complex Morse) theory, following the notation of \cite{witten}. For simplicity of notation, let us consider a scalar field theory consisting of a collection of $n$ real degrees of freedom $\{x_i\}_{i=1\cdots n}$ defined on a certain domain $\mcal{C}$ (we will take $\mcal{C}=\mbb{R}^n$). The path-integral expression for the expectation value of an observable $O$ is

\beq
\langle O \rangle=Z^{-1}\int\limits_\mcal{C}\mrm{d}^n x\,O\l[x\r]e^{-S\l[x\r]}\label{eq:morse_integral_O}
\eeq

with the partition function $Z$ given by the same integral without $O\l[x\r]$. The (complex) action $S=S_R+i\,S_I$ and the observable $O$ are assumed to be some holomorphic functions of the fields. Now we complexify the fields, letting $x_i\mapsto z_i=x_i+i\,y_i\in\mbb{C}$; Morse theory states that the following decomposition holds

\beq
Z=\sum_\sigma m_\sigma\int\limits_{\mcal{J}_\sigma} \mrm{d}^n z\,e^{-S\left(z\right)}=\sum_\sigma m_\sigma\,e^{-i\,S_I\l(z_\sigma\r)} \int\limits_{\mcal{J}_\sigma} \mrm{d}^n z\,e^{-S_R\left(z\right)}\label{eq:morse_integral_Z}
\eeq

 where $\sigma$ labels each critical point $z_\sigma$ of the action $S\l(z\r)$ in the complexified domain and $\mcal{J}_\sigma\subset\mcal{C}^n$ is the (\emph{stable}) Lefschetz thimble associated with the critical point $z_\sigma$. $m_\sigma\in\mbb{Z}$ are integer coefficients whose computation is potentially non-trivial (they will be discussed later). An analogous result holds for the integral in eq. (\ref{eq:morse_integral_O}). As previously stated, $S_I$ is constant along the thimble and therefore can be computed once and for all at the critical point: all is left is a real weight $e^{-S_R}$ which can be used for Monte Carlo simulations.

We now give the definition of stable thimble: it is a manifold of real dimension $n$ consisting of the union of all those curves of steepest-ascent (SA) for the ``action'' $S_R$ which fall into $z_\sigma$ at $\tau\rightarrow-\infty$, that is solutions of

\beq
\frac{\mrm{d}x_i}{\mrm{d}\tau} = \frac{\de S_R\l(x,y\r)}{\de x_i}\qquad\frac{\mrm{d}y_i}{\mrm{d}\tau} = \frac{\de S_R\l(x,y\r)}{\de y_i}\label{eq:SA_eqs}
\eeq

It is easy to check that, starting from the critical point at $\tau\rightarrow-\infty$, along these SA curves $S_R$ is always increasing, while $S_I$ stays constant.

As for the coefficients $m_\sigma$, they are defined as the intersection numbers between $\mcal{K}_\sigma$ and the original domain of integration $\mcal{C}$: $m_\sigma=\langle\mcal{C},\mcal{K}_\sigma\rangle$, with $\mcal{K}_\sigma$ the \emph{unstable} thimble associated with $z_\sigma$. The unstable thimble is given by the union of all the curves $z\l(\tau\r)$ that are solutions to (\ref{eq:SA_eqs}) which fall into $z_\sigma$ for $\tau\rightarrow+\infty$. The problem of knowing which thimbles contribute to the decomposition (\ref{eq:morse_integral_Z}) is a highly non trivial one (see \cite{originalPRD} for a detailed discussion). In the present work, we have analytical results available and so will discuss the relevance of different thimbles a-posteriori.

Let us now consider the integration measure appearing in (\ref{eq:morse_integral_Z}). The canonical basis of $\mbb{C}^n$ whose duals appear in $\mrm{d}^nz=\mrm{d}z_1\wedge\cdots\wedge\mrm{d}z_n$ is in general not parallel to the tangent space basis of the thimble at a generic point $z$ (we will name such tangent space $T_z\mcal{J}_\sigma$). Let $\{U^{\l(i\r)}\}_{i=1\cdots n}$ be the tangent space basis at $z$ (these are $n$ complex vectors with $n$ components forming the columns of an $n\times n$ unitary matrix $U$). Now we make a change of coordinates from the canonical ones of $\mbb{C}^n$ to the ones of $T_z\mcal{J}_\sigma$ (these are $n$ \emph{real} numbers $\{y_i\}_{i=1\cdots n}$). The integration measure thus becomes $\mrm{d}^nz=\det U\,\mrm{d}^ny$, where the term $\det U=e^{i\,\omega}$ is what is referred to as the \emph{residual phase}, which accounts for the non-parallelism of $T_z\mcal{J}_\sigma$ with the canonical basis of $\mbb{C}^n$ \cite{originalPRD,kiku,resphase,ournew}. This is a complex term which must be taken into account by using reweighting; this could in principle create a residual sign problem. We expect this not to be the case, as $\omega$ varies smoothly along SA curves. Moreover the regions in which it could get large are likely to be far from the critical point, that is regions where the action $S_R$ is large. However, this is something which should always be checked and so far the results have been very encouraging (see \cite{kiku} and \cite{resphase} for a more thorough discussion). In applying the thimble approach to the CRM theory, we also found that the residual phase can be safely taken into account with reweighting.

\section{An algorithm to sample on the thimble}

In this section a new algorithm to sample field configurations on a thimble is sketched (this algorithm was first proposed in \cite{franztalk14}), the interested reader can find all the details in \cite{ournew}.

We now make a change of notation and work with $2n$ real fields $\{\phi_i\}$ which consist in both the real and the imaginary parts of $z$. Let us begin by describing the tangent space to $\mcal{J}_\sigma$ in the vicinity of the critical point $\phi_\sigma$. In this region, we can expand the action

\beq
S^R\l(\phi\r)=S^R\l(\phi_\sigma\r)+\frac{1}{2}\Phi^TH\Phi+\mcal{O}\l(\phi^3\r)
\eeq

where the components of the $2n$-dimensional real vector $\Phi$ are $\Phi_i=\phi_i-\phi_{\sigma,i}$ and $H$ is the $2n\times 2n$ hessian matrix of $S_R$ evaluated at the critical point. The hessian can be put into diagonal form by the transformation $H=W\Lambda W^T$, with $W$ an orthogonal matrix whose columns are the eigenvectors of $H$. Because of holomorphicity of the action, the spectrum features $n$ positive eigenvalues $\{\lambda_i\}_{i=1\cdots n}$ and their negative counterparts (equal in modulus). The positive eigenvalues correspond to eigenvectors which span the tangent space to the stable thimble at the critical point: any combination of these gives a direction along which $S_R$ grows (we shall label this set of eigenvectors $\{v^{\l(i\r)}\}_{i=1\cdots n}$). A particular SA curve on the thimble can be identified defining the direction along which one leaves the critical point; this direction is given by an $n$-dimensional normalised\footnote{Being normalised, the vector $\hat{n}$ effectively encodes $n-1$ degrees of freedom, which, along with the time $t$ of integration of SA equations, is precisely what is needed to recover the original $n$ degrees of freedom.} vector $\hat{n}$. Therefore a point $\Phi\in\mcal{J}_\sigma$ can be unambiguously identified by a choice of $\hat{n}$ and the time $t$ of integration of the SA equations (\ref{eq:SA_eqs}). The mapping is thus

\beq
S^{n-1}\times\mbb{R}\ni\l(\hat{n},t\r)\leftrightarrow\Phi\in\mcal{J}_\sigma
\eeq

If one explores whole SA curves by integrating in $t\in\mbb{R}$, the thimble can then be sampled by picking up \emph{all} the possible SA curves. This is the essence of the algorithm which is described in this work.

In the following we need to compute the tangent space to the thimble at a generic point $\Phi$, for which we lack a local description. We shall resort to parallel-transporting the well-known tangent space at the critical point along the SA curve to the point $\Phi$. This is accomplished by solving a parallel-transport equation for each basis vector (\cite{originalPRD,kiku,resphase})

\beq
\frac{\mrm{d}V^{\l(i\r)}_j}{\mrm{d}t}=\sum_{k=1}^{2n}\frac{\de^2 S^R}{\de\phi_j\de\phi_k}V^{\l(i\r)}_k\qquad i=1\cdots n\quad j=1\cdots 2n\label{eq:PT_eqs}
\eeq

The asymptotic ($t\rightarrow-\infty$) solution to (\ref{eq:SA_eqs}) and (\ref{eq:PT_eqs}) is given by

\beq
t\ll 1
\bcas
\phi_j\l(t\r)\approx\phi_{\sigma,j}+\sum\limits_{i=1}^n v^{\l(i\r)}_j e^{\lambda_i t}n_i & j=1\cdots 2n\quad\l|\vec{n}\r|^2=1\\
V^{\l(i\r)}_j\l(t\r)\approx v^{\l(i\r)}_j e^{\lambda_i t} & j=1\cdots 2n\quad i=1\cdots n
\ecas
\label{eq:gauss_SA_PT}
\eeq

If we know a reference time $t_0\ll 1$ at which (\ref{eq:gauss_SA_PT}) holds, the latter can be used to provide an initial condition for a SA on the thimble.

We now rewrite the partition functions as a sum of contributions from all the possibile SA curves. We will take a few shortcuts (further details can be found in \cite{ournew}). The partition function function can be cast into the form

\beq
Z=\sum_\sigma m_\sigma\,e^{-i\,S_I\l(\phi_\sigma\r)}\int\limits_{\mcal{J}_\sigma}\prod\limits_{i=1}^n\mrm{d}y_i\,e^{-S_R}e^{i\,\omega}=\sum_\sigma m_\sigma\,e^{-i\,S_I\l(\phi_\sigma\r)}\int\prod_{i=1}^n\mrm{d}n_i\,\delta\l(\l|\vec{n}\r|^2-1\r)Z_{\hat{n}}\langle e^{i\,\omega}\rangle_{\hat{n}}
\eeq

by employing a Faddeev-Popov-like trick. The quantity $\langle f\rangle_{\hat{n}}$ may be regarded as the ``expectation value'' of $f$ on a \emph{single} SA curve

\beq
\langle f\rangle_{\hat{n}}=\frac{\int\limits_{-\infty}^{+\infty}\mrm{d}t\,\Delta_{\hat{n}}\l(t\r)e^{-S_R\l(\hat{n},t\r)}f\l(t\r)}{\int\limits_{-\infty}^{+\infty}\mrm{d}t\,\Delta_{\hat{n}}\l(t\r)e^{-S_R\l(\hat{n},t\r)}}
\label{eq:expect_f_n}
\eeq

where the quantity in the denominator can be regarded as a ``partial'' partition function $Z_{\hat{n}}$. An expression for $\Delta_{\hat{n}}\l(t\r)$ is worked out in \cite{ournew} in terms of the local (parallel-transported) basis vectors. The same vectors are then used to compute the complex $n\times n$ matrix $U$ whose determinant is the residual phase. The expectation value of an observable $O$ can be computed as

\beq
\langle O \rangle=\frac{1}{Z}\sum_\sigma m_\sigma\,e^{-i\,S_I\l(\phi_\sigma\r)}\int\prod_{i=1}^n\mrm{d}n_i\,\delta\l(\l|\vec{n}\r|^2-1\r)Z_{\hat{n}}\langle O\,e^{i\,\omega}\rangle_{\hat{n}}
\eeq

The $t$ integration which is evident in (\ref{eq:expect_f_n}) is carried out numerically while integrating (\ref{eq:SA_eqs}). We are then left with integration in $\hat{n}$-space. As each SA contributes to the total partition function with a weight $Z_{\hat{n}}/Z$, it would be reasonable to do some kind of importance sampling with respect to this measure. This is difficult, since computing $Z_{\hat{n}}$ requires integration along all the SA curve identified by $\hat{n}$. As a first try for the algorithm we have therefore used a static, crude Monte Carlo, that is we sampled the $\hat{n}$-space uniformly. This method can easily become very inefficient (\emph{e.g.} for large systems) and is the cause of the larger error bars in Figure \ref{fig:final_plot}; anyway, this crude approach proved to be enough to show the effectiveness of the thimble regularization for the CRM model in the range of parameters we explored.

\section{The CRM theory}

The Chiral Random Matrix theory we refer to is described in \cite{kim1} and is related to QCD in the $\epsilon$-regime of Chiral Perturbation Theory. The sign problem comes from the presence of a chemical potential $\mu$. In \cite{kim1} complex Langevin was used to compute the expectation value of the chiral condensate for different values of the quark mass and failures occured at low masses, where the algorithm seemed to follow the predictions of the phase-quenched theory instead of the exact one. In \cite{kim2} a different parametrization was proposed which made complex Langevin work fine. In this work we make use of the first parametrization and show that the thimble approach has no problems in computing expectation values of the condensate. The partition function of the CRM model is

\beq
Z_{N}^{N_{f}}(m)=\int\mrm{d}\,\Phi\mrm{d}\Psi\,{\det}^{N_f}\l(D(\mu)+m\r)\exp\l(-N\,\mrm{Tr}[\Psi^{\dagger}\Psi+\Phi^{\dagger}\Phi]\r),
\eeq

with

\beq
D\l(\mu\r)+m=
\bmat
m & i\cosh\l(\mu\r)\Phi+\sinh\l(\mu\r)\Psi\\
i\cosh\l(\mu\r)\Phi^\dag+\sinh\l(\mu\r)\Psi^\dag & m
\emat
\eeq

$\Phi$ and $\Psi$ are two $N\times N$ complex matrices, that is $\Phi=a+i\,b$ and $\Psi=\alpha+i\,\beta$. We have therefore $4N^2$ degrees of freedom which will be complexified, the thimble being a manifold of real dimension $4N^2$ embedded in an euclidean space of dimension $8N^2$. The observable we will be concerned with is the chiral condensate defined as

\beq
\frac{1}{N}\langle\bar{\eta}\eta\rangle=\frac{1}{N}\frac{\de}{\de m}\log Z\label{eq:ccond}
\eeq

In Figure \ref{fig:final_plot} we see the results of the thimble simulation for the real part of the condensate at fixed $N_f=2$ and $\tilde{\mu}=\sqrt{N}\mu=2$ as a function of $\tilde{m}=Nm$ for $N=1,2,3,4$. The exact results are recovered within errors, even in regions where the sign problem is severe (the exact result is very different from the phase-quenched one). All the simulations were carried out considering only the thimble attached to the trivial vacuum $a=b=\alpha=\beta=0$. We found no evidence of the relevance of more than this thimble, which was quite unexpected, especially at low dimensions. As a check, we tried to identify other critical points: we found two classes of them (there is a symmetry involved - see \cite{ournew}) but they all had

\beq
S_R\l(\phi_\sigma\neq 0\r)<S_R\l(\phi_\sigma=0\r)=\underset{\phi\in\mbb{R}^{4\l(N\times N\r)}}\min S_R\l(\phi\r)
\eeq

and by Morse theory they are all irrelevant in the decomposition (\ref{eq:morse_integral_Z}) \cite{originalPRD,witten,kiku}.

\begin{figure}[ht]
\begin{center}
\includegraphics[height=7cm,clip=true]{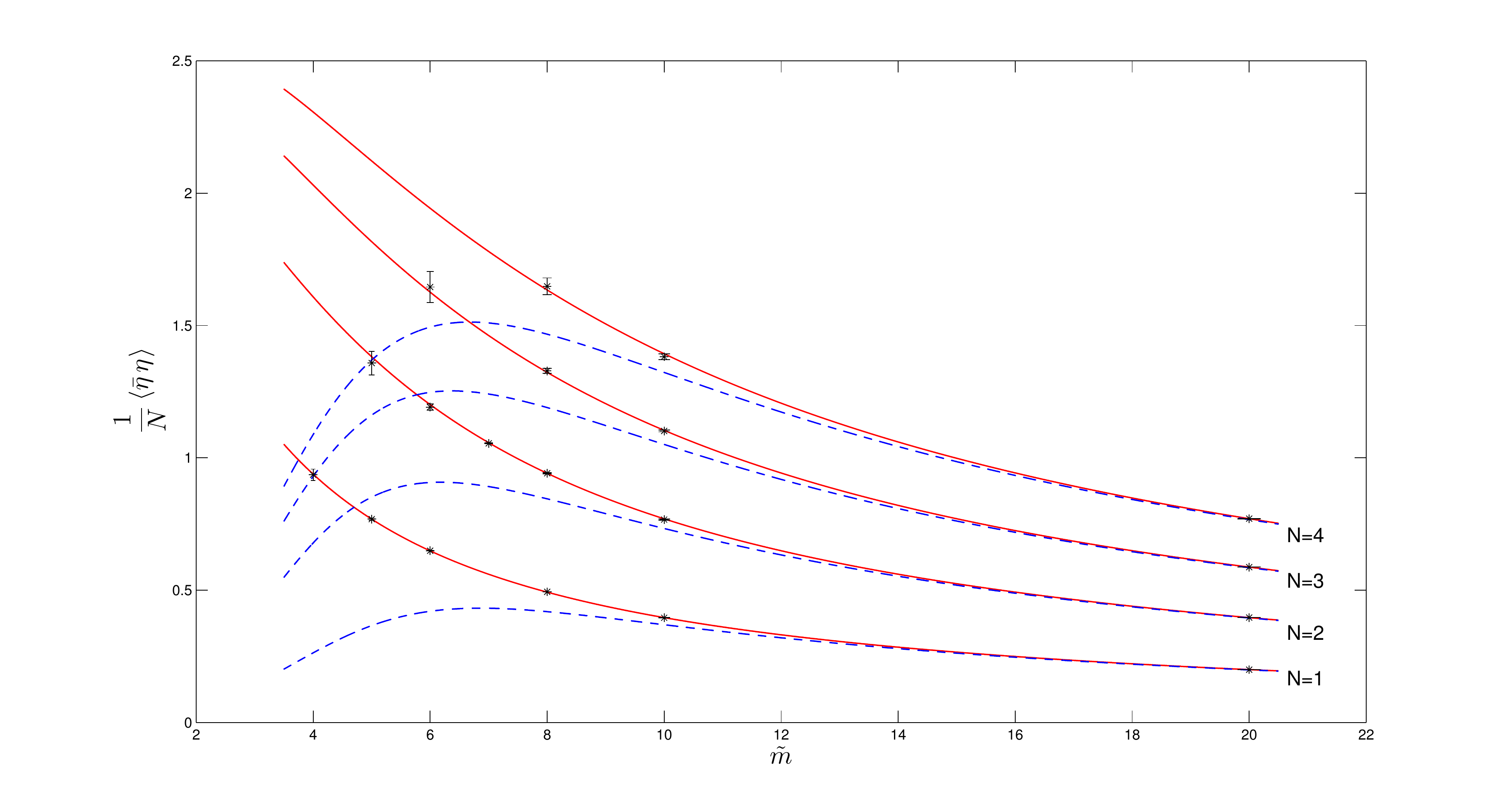}
\caption{Exact (solid red line), phase-quenched (dashed blue line) and thimble simulation results for the chiral condensate at fixed $N_f=2$, $\tilde{\mu}=2$.}
\label{fig:final_plot}
\end{center}
\end{figure}

\section{Gauge theories}

Gauge theories are the ultimate goal for the thimble regularization. Before tackling them, tests on low-dimensional toy models featuring gauge invariance are mandatory, being the approach still new and in need of some tuning. The details of the thimble formalism applied to theories featuring gauge invariance (as well as some examples) are discussed in \cite{franztalk15}. Here we just mention three fundamental steps.
The first step is the complexification of the gauge group (which we assume to be $\mrm{SU}\l(N\r)$)

\beq
\mrm{SU}\l(N\r)\ni U=e^{i x_aT^a}\rightarrow e^{i z_aT^a}=e^{i \l(x_a+i y_a\r)T^a}\in\mrm{SL}\l(N,\mbb{C}\r)
\eeq

then we have to rewrite the SA equations (\ref{eq:SA_eqs}) in a gauge-covariant form and the final ingredient is the parallel-transport equation for each basis vector, generalized to a non-abelian gauge group. For further details, the reader may consult \cite{franztalk15}.

\section{Conclusions}

We have discussed the thimble approach applied to the Chiral Random Matrix theory, yielding correct results in a region where the sign problem is quite severe. On this occasion we also tested for the first time on a multi-dimensional model the algorithm initially proposed in \cite{franztalk14}. The whole method proved to be quite effective, even in its crudest implementation, that is the static Monte Carlo for the extraction of the SA curves in $\hat{n}$-space. Moreover, we found no evidence for the presence of relevant thimbles beyond the one associated to the trivial vacuum (at least, in the region of parameter space we explored). This is of course something which is hard to generalise to a realistic field theory, nevertheless these first results are very encouraging with respect to more realistic applications of the thimble regularisation.

\section*{Acknowledgements}

We deeply thank L. Scorzato for the many useful and stimulating discussions. We also thank C. Torrero, K. Splittorff, Y. Kikukawa and H. Fujii.

\end{document}